\documentclass{elsart}

\usepackage{graphicx}
\usepackage{amssymb}

\begin{document}

\begin{frontmatter}



\title{Electron-nuclear entanglement in the cold lithium gas}


\author{Guo-Qiang Zhu, Jun-Wen Mao, You-Quan Li}

\address{Zhejiang Institute of Modern Physics, Zhejiang University, Hangzhou 310027,
P. R. China}

\begin{abstract}
We study the ground-state entanglement and thermal entanglement in the hyperfine
interaction of the lithium atom. We give the relationship between the entanglement and
both temperature and external magnetic fields.
\end{abstract}

\end{frontmatter}

% main text
\section{Introduction}
Quantum entanglement is an important prediction of quantum mechanics and constitutes
indeed a valuable resource in quantum information processing. In recent years, many
important results have been obtained in both experimental and theoretical
aspects\cite{nielsen}.

The effects of the temperature were taken into account and some authors used thermal
entanglement to study the electron spin system or other models at finite temperature in
Refs.\cite{arnesen,wang,sun}.  For example, X.G. Wang investigated the isotropic
Heisenberg XXX model at finite temperature\cite{wang}. Y. Sun et al. investigated the
thermal entanglement in the two-qubit Heisenberg XY model in nonuniform magnetic
fields\cite{sun}. On the other hand, the entanglement at the critical point is also a
hot point\cite{osterloh,osborne}. For instance, Osterloh et al. demonstrated that for a
class of one-dimensional magnetic systems, that entanglement shows scaling behavior in
the vicinity of the transition point\cite{osterloh}

In this paper, we study the entanglement in the $^6Li$ atom. Near absolute zero, the
atom will show some particular properties. As we know, in $^6Li$ atom the 3 electrons
possess spin $\frac{1}{2}$ and the nucleus has 3 protons and 3 neutrons so that the
total nuclear spin is $1$. Because the number of neutrons is odd, the atom obeys fermion
statistics. As we know, two research groups have succeeded in producing a Bose-Einstein
condensation of molecules made from pairs of fermion atoms\cite{jochim,regal}. Note that
the atoms are fermions but if considered as pairs they are bosons and therefore able to
condense in Bose-Einstein fashion. So it is necessary to study the entanglement
properties of the atom in the external fields at very low temperature.

In the following, the ground state entanglement of such a system is evaluated by means
of the method of Rungta's concurrence. Then we use negativity to study the entanglement
between the electrons and the nucleus at finite temperature.

\section{The model and Hamiltonian}
In the $^6Li$ atom, the electron spins are coupled to the nuclear spin by the hyperfine
interaction. The hyperfine line for the lithium atom has a measured magnitude of 228MHz
in frequency. Some calculation on the basis of first-order perturbation for the magnetic
dipole interaction between the electron and the nucleus gives contribution to the
coupling strength of $\mathbf{I\cdot S}$ term. In this paper, we study a bipartite
system, one is the nucleus which has the spin-$1$ and the other is the electron which
has total spin-$1/2$. The both parts are interacted with different external magnetic
fields, $B_{1}$ and $B_{2}$. Since the electrons have no orbital angular
momentum($L=0$), there is no magnetic field at the nucleus due to the orbital motion.
The Hamiltonian can be written as follows,
\begin{equation}
H=J(I_{x}\cdot S_{x}+I_{y}\cdot S_{y}+I_{z}\cdot S_{z})+C S_{z}+D I_{z},
\label{hamiltonian}
\end{equation}
where J is the coupling constant. Throughout the paper, the constant is set to unit.
$I_i$ is the nuclear spin which has spin-1 and $S_i$ is electron spin. The two
parameters are related to the external fields. They are given by
\begin{equation}
C=g\mu_{B}B_{1},\ \ D=-\frac{\mu}{I}B_{2}.
\end{equation}
In the Hamiltonian (\ref{hamiltonian}) the electronic orbital angular momentum $L$ has
been assumed to be zero. For the lithium atom, the nuclear magnetic moment
$\mu=0.822\mu_N$, where $\mu_{N}=e\hbar/(2m_{p})$. Since $|C/D|\sim m_{p}/m_{e}\approx
2000$, for most applications D can be neglected. At the same level of approximation the
$g$ factor of the electron may be put equal to 2\cite{pethick}. In this paper, the
spin-1/2 state has the bases as $|\uparrow\rangle$, $|\downarrow\rangle$ and the spin-1
state has the bases as $|\Uparrow\rangle$, $|0\rangle$, $|\Downarrow\rangle$. In the
above bases, the Hamiltonian can be rewritten as follows,
\begin{equation}
\left(%
\begin{array}{cccccc}
  \frac{1}{2}+\frac{C}{2}+D & 0 & 0 & 0 & 0 & 0 \\
  0 & -\frac{1}{2}-\frac{C}{2}+D & \frac{1}{\sqrt{2}} & 0 & 0 & 0 \\
  0 & \frac{1}{\sqrt{2}} & \frac{C}{2} & 0 & 0 & 0 \\
  0 & 0 & 0 & -\frac{C}{2} & \frac{1}{\sqrt{2}} & 0 \\
  0 & 0 & 0 & \frac{1}{\sqrt{2}} & -\frac{1}{2}+\frac{C}{2}-D & 0 \\
  0 & 0 & 0 & 0 & 0 & \frac{1}{2}-\frac{C}{2}-D \\
\end{array}%
\right).
\end{equation}
From the Hamiltonian one can easily obtain the eigenvalues and eigenvectors. During the
evaluation of the entanglement of the system, we will have to deal with the
high-dimensional Hilbert space. Rungta et al. raised a quantity they called it \emph{I
concurrence} to measure the high-dimensional bipartite pure state\cite{rungta}. The
quantity is given by
\begin{equation}
\mathfrak{C}(\rho)=\sqrt{2 \nu_{D_1} \nu_{D_2}(1-tr\rho_A^2)},
\end{equation}
where $D_1$ and $D_2$ are the dimensions of the Hilbert spaces and $\nu_{D_1},\nu_{D_2}$
are two parameter related to dimension. Here they can be set to unit. $\rho_{A}$ is the
reduced density matrix. The concurrence vanishes for unentangled state. Let
$D=\min\{D_1,D_2\}$. So the concurrence arranges from 0 to $\sqrt{2(D-1)/D}$.
\section{Entanglement in the presence of non-uniform fields}
In this section, we will study the ground-state entanglement of the system. Since the
parameter $C$ is much larger than $D$, D will be neglected.

(a)\ \ When $C<0$, the ground state energy is $(-1-\sqrt{9-4C+4C^2})/4$ and the
eigenvector is given by
\begin{equation}
|\Psi_1\rangle=\frac{1}{N_1}\left(\frac{1-2C-\sqrt{9-4C+4C^2}}{2\sqrt{2}}|0\downarrow\rangle+|\Downarrow\uparrow\rangle\right),
\end{equation}
where $N_1$ is the normalization factor. One can use Rungta's concurrence to measure the
entanglement and obtain
\begin{equation}
\mathfrak{C}(|\Psi_1\rangle)=\frac{4\sqrt{2}|1-2C-\sqrt{9-4C+4C^2}|}{(1-2C-\sqrt{9-4C+4C^2})^2+8}.\label{a}
\end{equation}

(b)\ \ When $C>0$, the ground state is $(-1-\sqrt{9+4C+4C^2})/4$ and the eigenvector is
given by
\begin{equation}
|\Psi_2\rangle=\frac{1}{N_2}\left(-\frac{1+2C+\sqrt{9+4C+4C^2}}{2\sqrt{2}}|\Uparrow\downarrow\rangle+|0\uparrow\rangle\right),
\end{equation}
where $N_2$ is the normalization factor. The concurrence is
\begin{equation}
\mathfrak{C}(|\Psi_2\rangle)=\frac{4\sqrt{2}|1+2C+\sqrt{9+4C+4C^2}|}{(1+2C+\sqrt{9+4C+4C^2})^2+8}.\label{b}
\end{equation}

\begin{figure}[h]
\begin{center}
\includegraphics[width=8 cm]{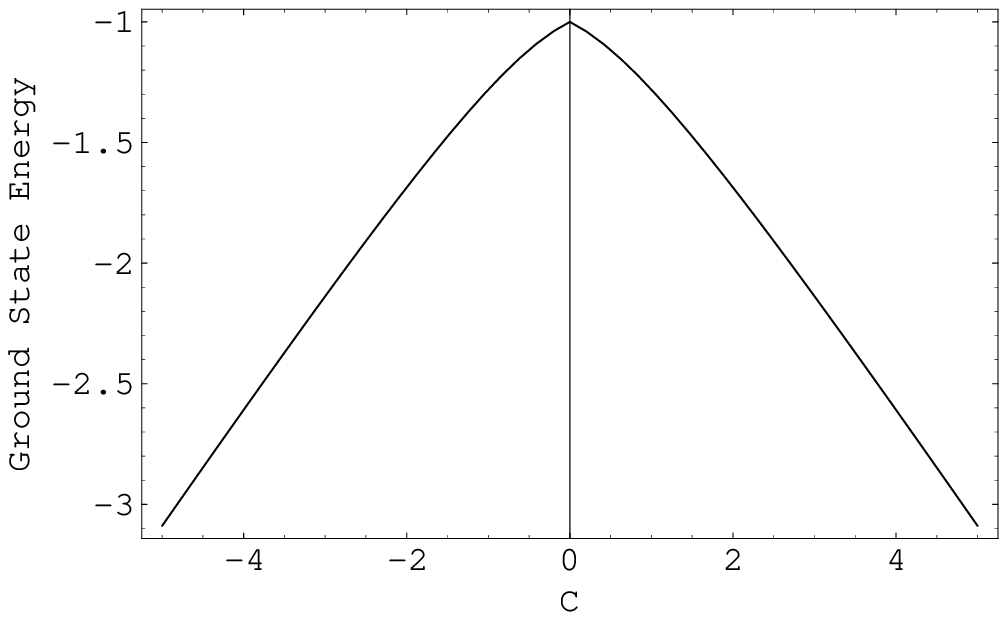}
\caption{\label{con} Ground state energy when $D=0$}
\includegraphics[width=8 cm]{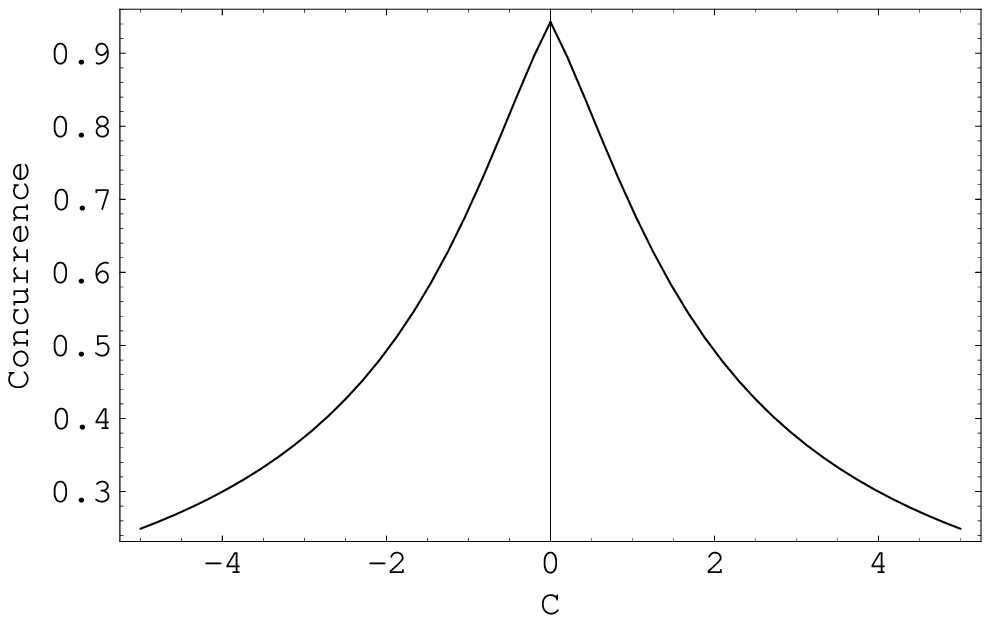}
\caption{\label{concurrence}Ground-state entanglement versus $C(=g\mu_{B}B)$.}
\end{center}
\end{figure}

The above results can be plotted in Fig.\ref{con} and Fig.\ref{concurrence}.
Fig.\ref{con} shows the relationship between the ground-state energy and external
magnetic fields. Fig.\ref{concurrence} shows the relationship between the ground-state
entanglement and external magnetic fields C.
From the Fig.\ref{concurrence}, one can see when the parameter $C$ approaches zero,
i.e., the magnetic fields is vanishing, the concurrence approaches its maximum
$2\sqrt{2}/3\approx 0.943.$ In fact, when the magnetic field is absent, the ground state
is degenerate, it will be discussed in the following paragraph.

(c)\ \ When $C=0$, i.e. the magnetic field is absent, the ground state will be double
degenerate. As we know, if the ground state is degenerate the zero-temperature ensemble
becomes an equal mixture of all the possible ground states\cite{osborne}.  In this case,
the thermal ground state may be written
\begin{eqnarray}
\rho=\frac{1}{2}|\phi_1\rangle\langle\phi_1|+\frac{1}{2}|\phi_2\rangle\langle\phi_2|,
\label{c}
\end{eqnarray}
where
$$|\phi_1\rangle=-\sqrt{\frac{1}{3}}|0\downarrow\rangle+\sqrt{\frac{2}{3}}|\Downarrow\uparrow\rangle,
\ \
|\phi_2\rangle=-\sqrt{\frac{2}{3}}|\Uparrow\downarrow\rangle+\sqrt{\frac{1}{3}}|0\uparrow\rangle.$$
One can use \emph{Negativity} to measure the entanglement of the state. The Negativity
was introduced by G. Vidal \emph{et al}\cite{vidal}. The  quantity is given by
\begin{equation}
\mathfrak{N}(\rho)\equiv\frac{\|\rho^{T_{A}}\|_{1}-1}{2},
\end{equation}
where the trace norm is defined by $\|X\|_{1} \equiv tr\sqrt{X^{\dagger}X}$ and $T_{A}$
denotes the partial transpose of the bipartite mixed state $\rho$. Negativity vanishes
for unentangled states. It is easy to know that Negativity of the state (\ref{c}) is
$1/3$.

\section{Thermal entanglement}
As we know, since thermal entanglement was introduced in 1998\cite{arnesen}, many
efforts are devoted to study the thermal state of qubit system. Here we want to use
Negativity to study the thermal entanglement of mixed-spin bipartite system at finite
temperature.

\begin{figure}
\begin{center}
\includegraphics[width=10 cm]{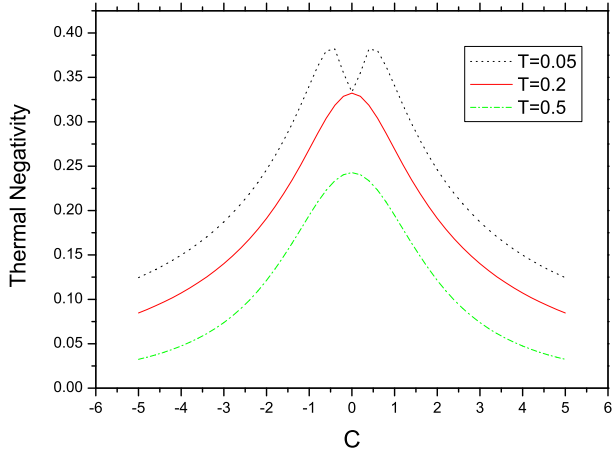}
\caption{\label{thermal} Thermal negativity versus $C(=g\mu_{B}B)$ at various
temperatrues.}
\end{center}
\end{figure}

Here $\rho$ stands for the Gibbs density operator, $\rho=\frac{1}{Z} \exp(-H/kT)$, where
$Z=tr\exp(-H/kT)$ is the partition function, H is the Hamiltonian, T is temperature and
$k$ is Boltzmann's constant which we henceforth will set equal to unit. In the bases
$|\Uparrow\uparrow\rangle$, $|\Uparrow\downarrow\rangle$, $|0\uparrow\rangle$,
$|0\downarrow\rangle$, $|\Downarrow\uparrow\rangle$, $|\Downarrow\downarrow\rangle$, the
thermal state $\rho$ can be rewritten as follows,
\begin{equation}
\left(%
\begin{array}{cccccc}
  \omega_{1} & 0 & 0 & 0 & 0 & 0 \\
  0 & \omega_{2} & s_{1} & 0 & 0 & 0 \\
  0 & s_{1} & \omega_{3} & 0 & 0 & 0 \\
  0 & 0 & 0 & \omega_{4} & s_{2} & 0 \\
  0 & 0 & 0 & s_{2} & \omega_{5} & 0 \\
  0 & 0 & 0 & 0 & 0 & \omega_{6} \\
\end{array}%
\right).
\end{equation}
One can use Negativity to study the thermal entanglement of the state. Here the
parameter $D$ is also neglected. At various temperatures, the Negativity versus $C$ can
be plotted in Fig.\ref{thermal}. From this, one can find when the temperature is low
enough, at the point $C=0$, the Negativity curve has a local minimum. However, as the
temperature increases and when it is larger than a critical value $T_{C}=0.107$, the
local minimum changes to be a local maximum. In detail, when $T=0.5$, at the point
$C=0$, the Negativity is 0.243. This is a local maximum. When $T=0.05$, the Negativity
is 0.333, but it is a local minimum. One can see that the temperature can efficiently
affect the entanglement of the system.
\section{Summary and acknowledgements}
In this paper, we studied the entanglement of the electron-nuclear entanglement of
$^6Li$ atom in presence of uniform external magnetic fields. We studied the ground-state
entanglement at zero temperature and thermal entanglement at finite temperature
respectively. We gave the relationship between the entanglement and the magnetic fields
or temperature. In order to detect the entanglement of the system, the surrounding
temperature should be near absolute zero.

This work was supported by NSFC No. 10225419 and 90103022.

\end{document}